\documentclass[aps,prb,amsmath,amssymb,superscriptaddress,twocolumn,color,epsfig,graphicx,bm,floatfix]{revtex4-1}
\usepackage{amsmath}    % need for subequations
\usepackage{mathtools}
\usepackage{siunitx}

\usepackage{tablefootnote}
\usepackage[caption=false]{subfig}
\usepackage{graphicx}   % need for figures
\usepackage{verbatim}   % useful for program listings
\usepackage{color}      % use if color is used in text
\usepackage{hyperref}   % use for hypertext links, including those to external documents and URLs
\usepackage{mhchem}  

\begin{document}

\author{Hossein Asnaashari Eivari}
\affiliation{Institute for Advanced Studies in Basic Sciences, P.O. Box 45195-1159, Zanjan, Iran}
\affiliation{University of Zabol, P.O. Box 98615-538, Zabol, Iran}
\author{S. Alireza Ghasemi}
\email{aghasemi@iasbs.ac.ir}
\author{Hossein Tahmasbi}
\author{Samare Rostami}
\author{Somayeh Faraji}
\author{Robabe Rasoulkhani}
\affiliation{Institute for Advanced Studies in Basic Sciences, P.O. Box 45195-1159, Zanjan, Iran}
\author{Stefan Goedecker}
\affiliation{Department of Physics, Universit\"{a}t Basel, Klingelbergstr. 82, 4056 Basel, Switzerland}
\author{Maximilian Amsler}
\email{amsler.max@gmail.com}
\affiliation{Department of Materials Science and Engineering, Northwestern University, Evanston, Illinois 60208, USA}

%\title{Low density quasi two dimensional \ce{TiO2} hexagonal structure}
\title{A two-dimensional hexagonal sheet of TiO$_2$}

\date{\today}

%----------------------------------------------------------------------------------------
\begin{abstract}
We report on the \textit{ab initio} discovery of a novel putative ground state for quasi two-dimensional \ce{TiO2} through a structural search using the minima hopping method with an artificial neural network potential. The structure is based on a honeycomb lattice and is energetically lower than the experimentally reported lepidocrocite sheet by 7~meV/atom, and merely 13~meV/atom higher in energy than the ground state rutile bulk structure. According to our calculations, the hexagonal sheet is stable against mechanical stress, it is chemically inert and can be deposited on various substrates without disrupting the structure.  Its properties differ significantly from all known \ce{TiO2} bulk phases with a large gap of 5.05~eV that can be tuned through strain engineering.
\end{abstract}

\maketitle

Low dimensional materials have attracted significant interest in the recent years, especially since the discovery of graphene in 2004 through its exfoliation from graphite.~\cite{Novoselov2004} Due to its extraordinary properties, ranging from high mechanical strength~\cite{lee_measurement_2008} and the presence of massless Dirac electrons~\cite{novoselov_two-dimensional_2005} to excellent thermal conductivity,~\cite{Xu2013} many other 2D materials have been considered and isolated for potential applications in electronics and energy conversion.~\cite{mas-balleste_2d_2011} Buckled graphene analogues like silicene~\cite{takeda_theoretical_1994,guzman-verri_electronic_2007,Cahangirov2009} have been grown on silver substrates,~\cite{vogt_silicene:_2012} and germanene~\cite{takeda_theoretical_1994} was synthesized on gold.~\cite{davila_germanene:_2014} Similarly, the theoretically predicted boron counterpart~\cite{amsler_conducting_2013,tang_novel_2007,tang_first-principles_2010,lu_binary_2013} with a partially filled honeycomb lattice was recently realized on silver substrates~\cite{mannix_synthesis_2015,zhang_substrate-induced_2016,Feng2016} and confirmed to exhibit Dirac fermions.~\cite{feng_dirac_2017} This class of materials however lacks a finite band gap, rendering them unsuitable for transistor electronics.

On the other hand, promising 2D semiconductors have been investigated for potential applications in nano electronics, optoelectronics and photonics.~\cite{ugeda_giant_2014,qian_quantum_2014,wang_electronics_2012,radisavljevic_mobility_2013,mak_tightly_2013} Besides the wide band gap h-BN monolayer,~\cite{Ivanovskii2012,Xu2013} transition metal di-chalcogenides (TDM)~\cite{Xu2013,Ivanovskii2012} have been intensely studied, which exhibit gaps that can be readily tuned based on the number of stacked monolayers,~\cite{Xu2013} and electronic properties that can be fundamentally modified through in-plane~\cite{duesberg_heterojunctions_2014,zhang_synthesis_2015,duan_lateral_2014,huang_lateral_2014,gong_vertical_2014} and vertical heterostructuring.~\cite{gong_vertical_2014,roy_graphene-mos2_2013,hong_ultrafast_2014,bhimanapati_recent_2015,xia_recent_2017} More recently, layered metal oxides have emerged as a new family of 2D materials, which exhibit an overall higher stability in air compared to TDMs.~\cite{geim_van_2013} Besides monolayers of \ce{MoO3},~\cite{du_iso-oriented_2016} \ce{WO3},~\cite{wang_preparation_2015} SnO,~\cite{saji_2d_2016} and perovskite type oxides,~\cite{mas-balleste_2d_2011}  \ce{TiO2} is considered one of the most promising candidates. As an important multi-functional oxide widely used in industrial products ranging from consumer goods to advanced materials for photovoltaic cells,~\cite{Grtzel2001} sensors,~\cite{Bai2014} hydrogen storage,~\cite{Chen2007} and rechargeable lithium ion batteries,~\cite{Kerisit2010,Zhang2013,Dylla2012_a,Dylla2012_b} titania based materials have also drawn the focus of research for nanoelectrics due to their structure dependent, high dielectric constant.~\cite{Osada2011} \ce{TiO2} nanosheets and nanostructures are touted to possess enhanced photocatalytic, photovoltaic, electrochemical and dielectric properties,~\cite{Leng2014,Chen2016,Wang2014,Lin2012} and immense efforts have been made to improve their performance by fabricating controlled size and/or shape at the nano scale.~\cite{Chen2007,Niu2014}

Despite these efforts, the atomistic structures in the majority of studies on 2D \ce{TiO2} remains unclear, and its ground state structure is not fully resolved. Only a few studies explicitly investigate the microscopic structure of \ce{TiO2}  nanosheets,~\cite{Bandura2014,Evarestov2010,Ferrari2010} and almost all theoretical studies are based on slabs cut out of bulk structures along different crystallographic planes. These include anatase (001)~\cite{Ferrari2010,Atrei2010} and (101),~\cite{Bandura2014,Evarestov2010,Vittadini2008} fluorite (111),~\cite{Niu2014,Bandura2014} rutile (011),~\cite{Niu2014}  (110)~\cite{Niu2014,Bandura2014,Yang2016} and (100),~\cite{Atrei2010} and \ce{TiO2}(B) (001).~\cite{Vittadini2010} The only fully reconstructed, truly 2D material of \ce{TiO2} known to date is the lepidocrocite type nanosheet (LNS),~\cite{Bandura2014,Vittadini2010,Atrei2010,Wang2014} which was also found to exhibit the lowest formation energy among all sheets reported so far in the literature. Using density functional theory (DFT) calculations, Orzali~\textit{et al.}~\cite{Orzali2006} showed that the LNS can be readily obtained from a slab of  anatase (001) with six atomic layer by a simple structural rearrangement.

While the practice of using bulk-like structures as a starting point for theoretical investigations is understandable, it is based on the assumption that atomically thin layers of \ce{TiO2} will adopt a structure close to the crystalline phase. However, physical epitaxy experiments have shown that many 2D materials often exhibit structural motifs that strongly differ from  any  bulk polymorphs.~\cite{Feng2016,Albert2009} To address this issue, we performed a global search for 2D \ce{TiO2} materials using a constrained minima hopping (MHM)~\cite{Goedecker2004,Amsler2010} approach together with a recently developed high dimensional artificial neural network potential based on the charge equilibrated neural network technique (CENT)~\cite{Ghasemi2015}. A plethora of structures were recovered, including the majority of the previously reported sheets discussed in the literature, indicating that the structural search was sufficiently converged. The most promising candidate structures were subsequently refined at the DFT level. 

\begin{figure}
\subfloat[\label{sfig:a}]{\includegraphics[width=0.52\columnwidth]{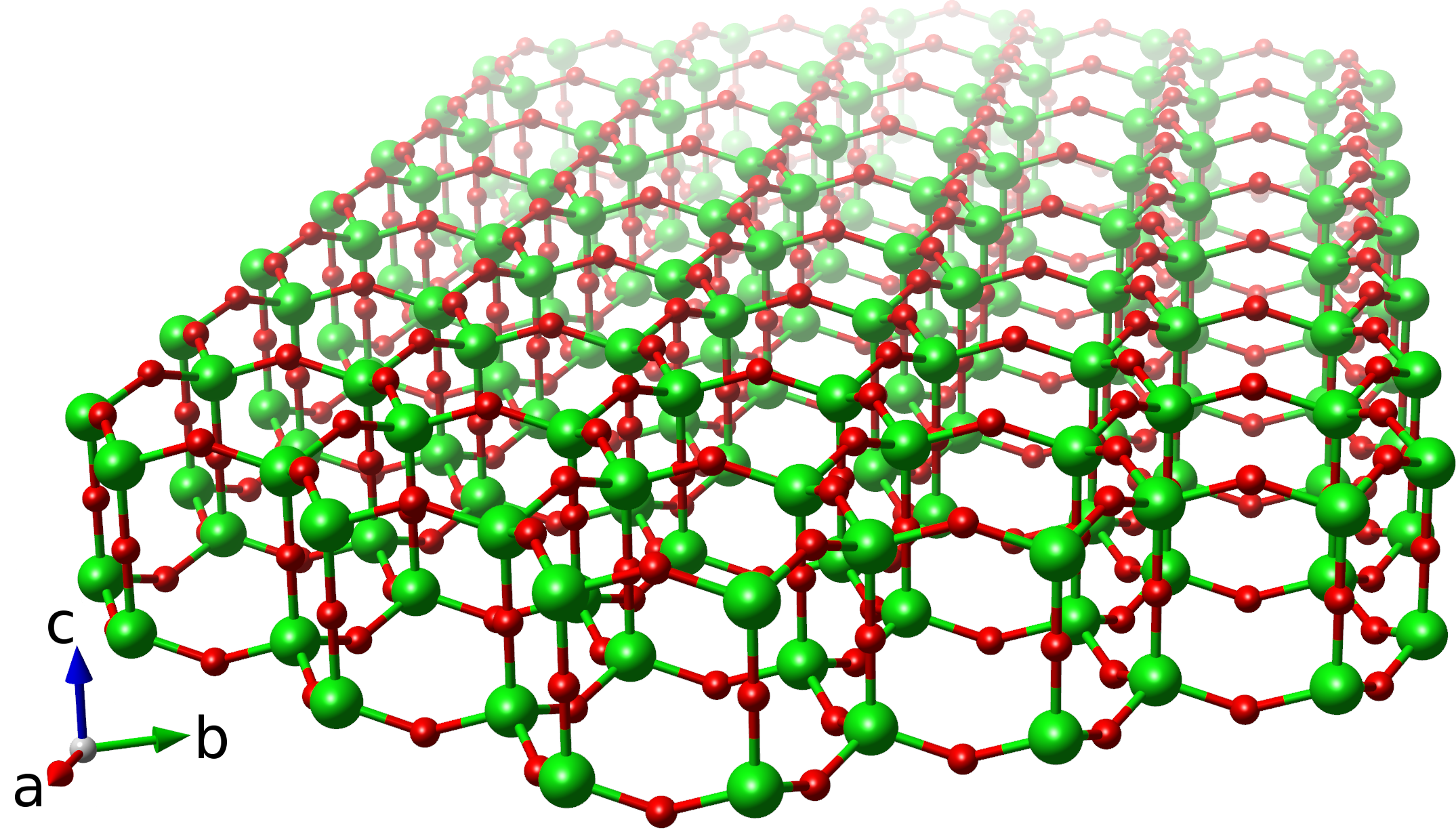}}\hfill
\subfloat[\label{sfig:b}]{\includegraphics[width=0.45\columnwidth]{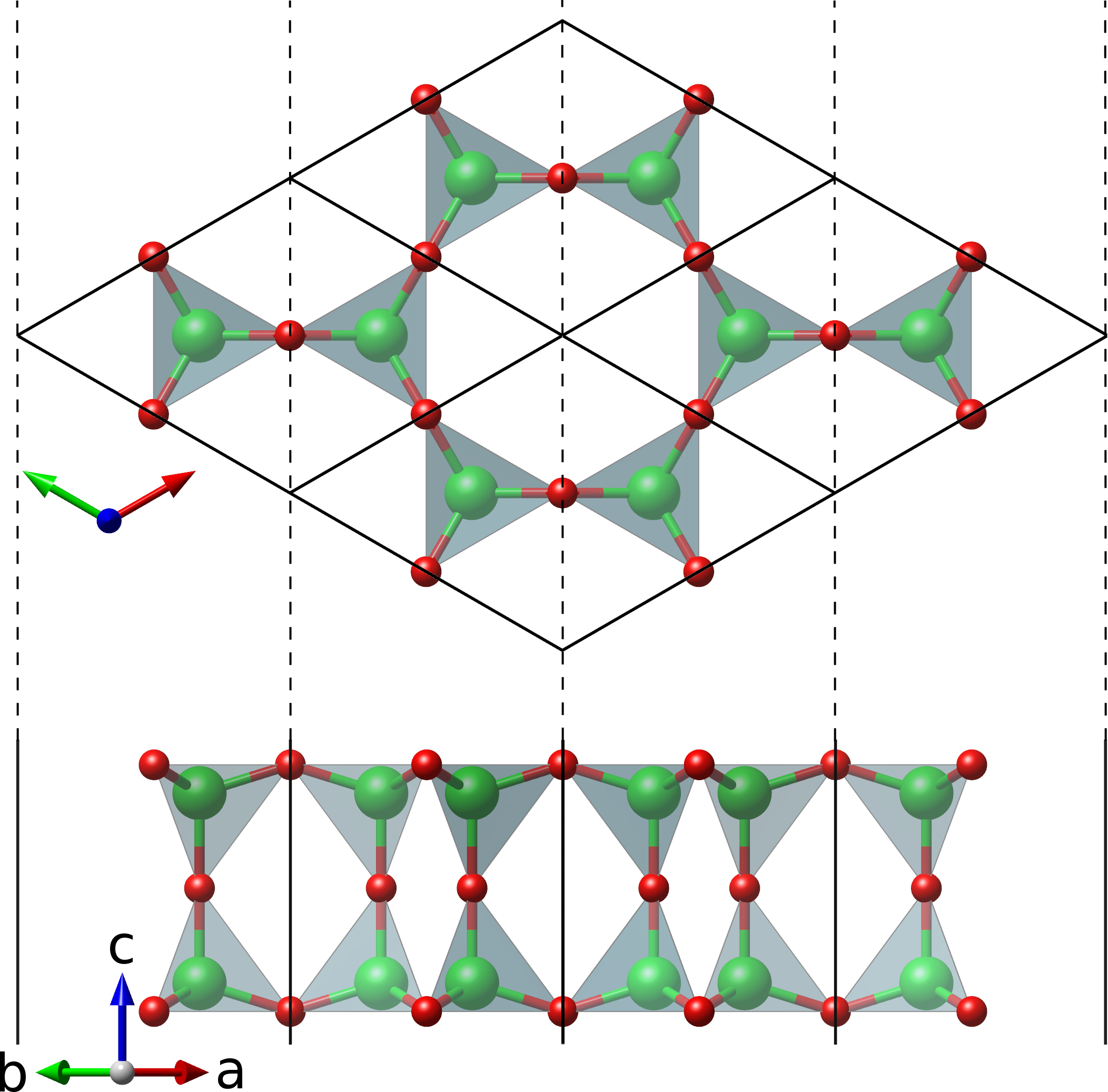}}

\caption{Panel (a) shows a single layer of the HNS from a perspective view. Panel (b) shows the projected views along the c direction and along the a-b plane, where the Ti-O bonds are represented by gray polyhedra. The green (large) and red (small) spheres represent the Ti and O atoms, respectively. 
\label{fig:191}}
\end{figure}

We discovered a novel nanosheet with hexagonal symmetry to have the lowest energy, hereafter referred to as hexagonal nano sheet (HNS), shown in panel (a) of Figure~\ref{fig:191}. In particular, the HNS is energetically favorable compared to the LNS, which had been considered to be the most stable 2D structure of \ce{TiO2} to date. From the structural point of view, the unit cell of HNS is composed of four-fold coordinated Ti and two-fold coordinated O atoms, arranged in a lattice with  $P6/mmm$ symmetry (see Table~\ref{table:wyckoff} for the lattice vectors and atomic coordinates). It is isostructural to a previously reported silicon dioxide sheet grown on a Ru surface.~\cite{Lffler2010} The \ce{TiO4} units form tetrahedra in two layers, which are stacked on top of each other and linked by shared corners, as illustrated in the side view in panel (b) of Figure~\ref{fig:191}. Large, hexagonal pores are formed along the out-of-plane axis and arranged in a honeycomb lattice, as shown from the top view in panel (b) of Figure~\ref{fig:191}. In strong contrast to the HNS, the LNS has a much denser structure, where each Ti atom is six-fold and the O atoms are two-fold or four-fold coordinated, respectively. The edge-sharing octahedra are arranged in an orthorhombic lattice, forming a sheet which is $0.4$~\AA\ thinner than the HNS. Similar to the HNS, many other 2D materials have a similar hexagonal lattice: the large lattice constant of 6.03~\AA\, in the HNS compares for example with values of 2.46~\AA\, in graphene,~\cite{Liu2007} 3.90~\AA\, in silicene~\cite{Cahangirov2009} and 3.15~\AA\,  in \ce{MoS2},~\cite{Dickinson1923} potentially allowing the construction of various (commensurate) 2D heterostructured materials with improved properties.

Due to the unique hexagonal nano pores in the HNS we expect it to have a very low atomic density compared to other sheets. To quantify the density we considered as the volume the product of the thickness $h$ of each sheet and the area $A$ of the unit cell along the two periodic directions, where $h$ is given by the distance between the two outmost atoms (column 4 in Table~\ref{tab:energies}). Furthermore, we analyzed the bond lengths in all 2D structures and crystalline polymorphs by comparing the minimum and maximum bonds in each system (column 5 and 6 in Table~\ref{tab:energies}). In general, small maximum bond length indicate strong and stiff bonds between Ti and O atoms. It is somewhat surprising that, although the HNS exhibits the shortest maximum bond lengths, its pseudo-density is the lowest among all sheets with a value of only 3.536~g/cm$^3$, 16~\% lower than any other sheet, indicating that the nano pores are stabilized through strong interatomic bonds. Furthermore, we also quantify the roughness of the surface in terms of the amount of buckling in the outermost atomic layers (SB), essentially corresponding to the out-of-plane distance of the two outmost Ti and O atoms (column 3 in Table~\ref{tab:energies}).

From the energetic point of view we consider the dissociation energies per atom as a measure of (meta) stability and compare them to the LNS and other configurations that have been reported in the literature (column 2 in Table~\ref{tab:energies}). The values were computed according to  $E_{\text{diss}}=(E-N_\text{Ti}E_\text{Ti}-N_\text{O}E_\text{O})/N$, where $N_\text{Ti}$ and $N_\text{O}$ are the number of Ti and O atoms in the cell, respectively, with the corresponding energies of the isolated atoms $E_\text{Ti}$ and $E_\text{O}$, whereas $E$ is the energy of the whole system and $N$ is the total number of atoms per cell. 
All structures that were directly cut from bulk have significantly higher $E_\text{diss}$ than the HNS, at least for the slab thicknesses that we considered in our study. In fact, the HNS is only $45$ and $13$~meV/atom higher in energy than bulk anatase and rutile, respectively, but most importantly $7$~meV/atom lower than the LNS, rendering it the most stable nanosheet among all known quasi-2D nano structures.
A comparative summary of the quantities discussed above is listed in Table~\ref{tab:energies}, showing that the HNS simultaneously has the lowest density and dissociation energy with respect to bulk anatase, $\Delta E_\text{diss}$.

\begin{table}[tb]
\centering
\caption{The structural parameters of the HNS with the Wyckoff positions of Ti and O. The values of $x$ and $y$ are given in reduced coordinates, whereas the $z$ is given in cartesian coordinates in units of~\AA. The equilibrium lattice parameter is $a=6.03$~\AA.)
\label{table:wyckoff}}
\small
\begin{ruledtabular}
\begin{tabular}{l@{}c*{3}{c}}
%\hline 
   &   & $x$ & $y$ & $z$ \\
[0.5ex]
\hline
Ti  &   4h &  $1/3$ & $2/3$ & $1.82$ \\
O1  &   6i &  $1/2$ & $0/0$ & $2.36$ \\
O2  &   2d &  $1/3$ & $2/3$ & $0.00$ \\
\end{tabular} 
\end{ruledtabular}  
\end{table}

\begin{center}
\begin{table}[!htbb]
    \caption{Table contains the dissociation energy with respect to anatase ($\Delta E_\text{diss}$) in meV/atom,
surface buckling (SB) in \AA, pseudo-density in g/cm$^3$,
maximum Ti-O bond length ($d_\text{min}$) and minimum Ti-O band
length ($d_\text{max}$) in \AA, and HSE06 level band-gap
energy ($E_g$) in eV. For structures cut from bulk material the last column indicates the number of atomic layers they contain as a measure of the initial slab thickness ($N_\text{layer}$).  Lattice parameters of rutile and anatase phases in our calculations are $a=4.65$~\AA~ and $c=2.97$~\AA~ for rutile and $a=3.80$~\AA~ and $c=9.70$~\AA~ for anatase.  These values compare well with experimental data, i.e.  $a=4.59$~\AA~ and $c=2.95$~\AA~ for rutile~\cite{Burdett1987} and $a=3.78$~\AA~ and $c=9.50$~\AA~ for anatase.~\cite{Burdett1987} The calculated lattice parameters of  LNS ($a=3.02$~\AA, $b=3.74$~\AA)  are  in good agreement with the theoretical values of Sato~\textit{et al.},~\cite{Sato2003} and close to the experimental values of Orzali~\textit{et al.}~\cite{Orzali2006} ($^a$represent bulk densities).
\label{tab:energies}}
\begin{ruledtabular}
\begin{tabular}{l@{}c*{6}{c}}
%\hline 
    structure          & $\Delta E_\text{diss}$ & SB   &den.& $d_\text{min}$ & $d_\text{max}$ & $E_g$  & $N_\text{layer}$\\ \hline
%[0.5ex]
rutile(100)                & $158$ & $>1.06 $  & $4.199$ & $1.83$ & $2.09$ & $4.84$ & $6$ \\
fluorite(111)              & $137$ & $>1.30$   & $5.967$ & $1.85$ & $2.07$ & $4.54$ & $3$ \\
anatase(101)               & $120$ & $>1.10$   & $5.967$ & $1.78$ & $2.10$ & $4.87$ & $6$ \\
rutile(110)                & $83$ & $>1.06$   & $4.597$ & $1.78$ & $2.14$ & $3.95$ & $6$ \\
\ce{TiO2}(B)(001)          & $60$ & $0.67$  & $4.951$ & $1.81$ & $2.07$ & $4.16$ & $7$ \\
LNS                        & $52$ & $1.03$    & $5.481$ & $1.83$ & $1.97$ & $4.65$ & $6$ \\
HNS                        & $45$ & $0.54$    & $3.536$ & $1.82$ & $1.82$ & $5.05$ &     \\
rutile                     & $32$ &           & $4.127^a$ & $1.99$ & $2.00$ & $3.40$ &     \\
anatase                    & $0$ &           & $3.765^a$ & $1.94$ & $2.00$ & $3.58$ &     \\

\end{tabular} 
% \footnotetext{bulk density}

\end{ruledtabular}  
\end{table}
\end{center}

We further studied the electronic properties of the new HNS  using hybrid functionals due to the well-known severe underestimation of the band gaps of semilocal functionals. The HSE06~\cite{Krukau2006} and PBE0~\cite{Paier2005} functionals were employed to assess their accuracy by computing the band gaps of the experimentally well studied rutile and anatase phases. The values for HSE06 with gaps of $3.40$~eV and $3.58$~eV for rutile and anatase, respectively, are closer to the experimental values ($3.0$~eV and $3.2$~eV for rutile and
for anatase, respectively~\cite{Burdett1987}) than PBE0, which significantly overestimates the gap with $4.16$~eV and $4.32$~eV for rutile and anatase, respectively. Hence, all subsequent electronic structure calculations were performed with HSE06, and the band gaps for the structures considered in this work are compiled in Table~\ref{tab:energies}.

\begin{figure}
\subfloat[\label{sfig:b}]{\includegraphics[width=1.0\columnwidth]{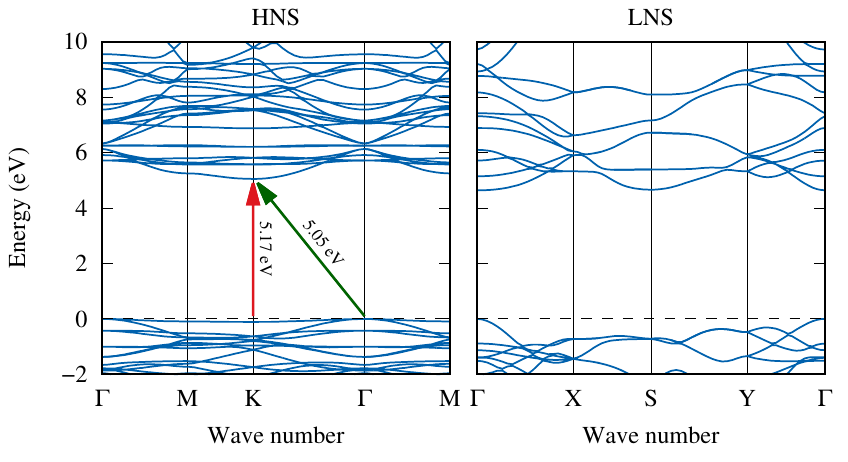}}\\
\subfloat[\label{sfig:a}]{\includegraphics[width=0.9\columnwidth]{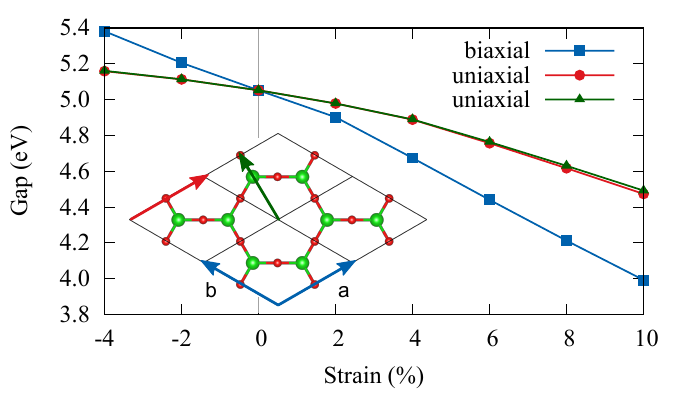}}
\caption{(a) Electronic band structures of HNS (left) and LNS (right) along a path in the 2D Brillouin zone with the HSE06 functional. The red and green arrow indicate the values of the direct and indirect band gap, respectively. (b) The band gap energies for different values of strain. The color coded arrows in the inset indicate the direction of biaxial and uniaxial strain.
\label{fig:eband_191_lepid}}
\end{figure}

\begin{figure}
\includegraphics[width=0.9\columnwidth]{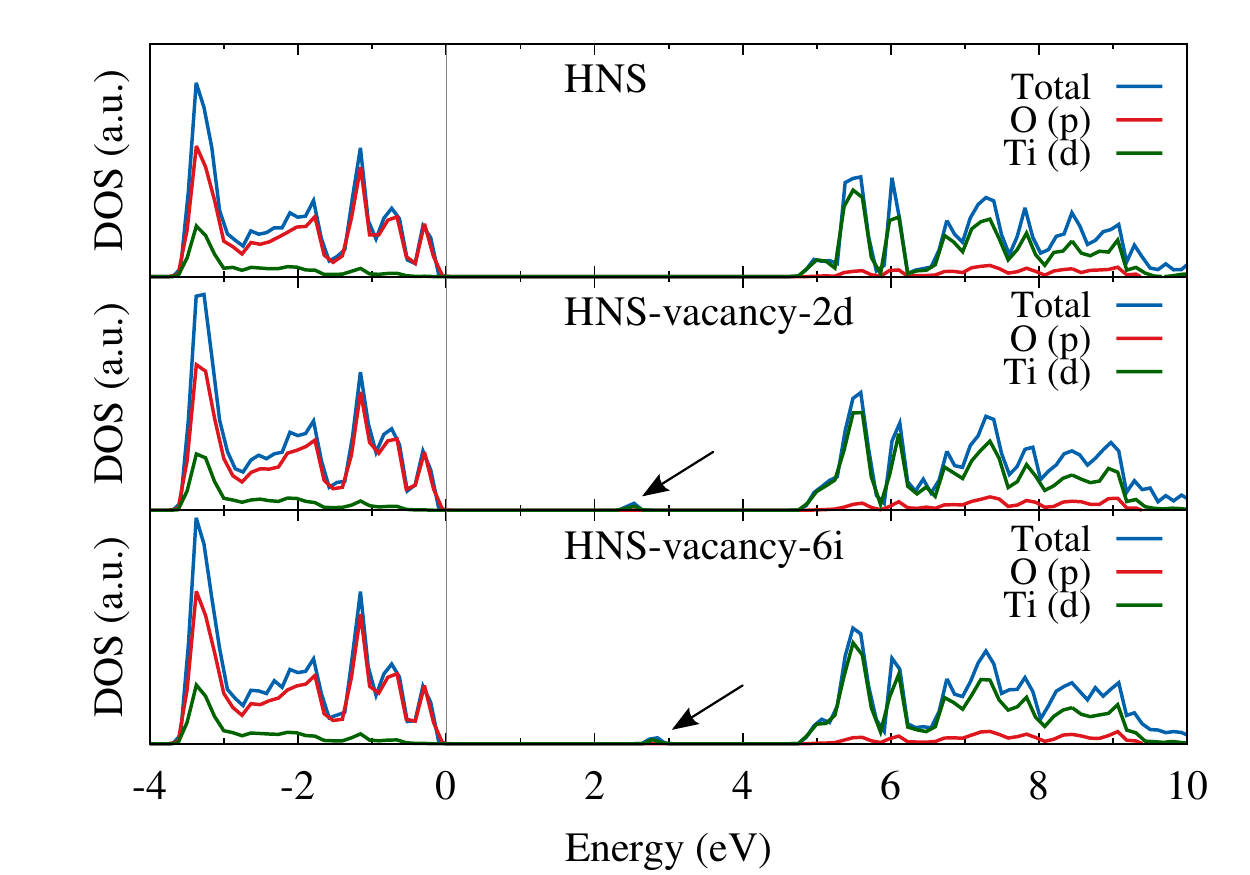}
\caption{Electronic pDOS of the pristine HNS (top panel) and the sheet with oxygen vacancies (two bottom panels) at the HSE06 level. The Fermi level of the two vacancy structures was shifted to align with the pristine HNS:  the defect states are occupied. 
\label{fig:DOS_hex_v1_v2_arrow}}
\end{figure}

The band structure for the pristine HNS and LNS are shown in Figure~\ref{fig:eband_191_lepid}, revealing that the LNS has a direct band gap, in contrast to the HNS. However, the valence band of the HNS exhibits very little dispersion, leading to a quasi-direct gap at the $K$-point with a value of 5.17~eV.
The partial DOS of the HNS, shown in the top panel of Figure~\ref{fig:DOS_hex_v1_v2_arrow}, indicates that the main contribution to the valence band stems from oxygen $p$-states, while empty $d$-states of the Ti atoms contribute mainly to the conduction bands. The magnitude of the HNS gap is very large, and in fact larger than any other \ce{TiO2} polymorph studied here, and lies well beyond the range of visible light. However, there are various methods to tune band gaps which could be employed in practice, such as impurity doping, introducing vacancies, elemental substitution, and creating heterostructures with other 2D building blocks.~\cite{Ramasubramaniam2011,Kanhere2015,Yun2017,Wang2014-a} Since defects can significantly influence the gap energies, the effect of oxygen vacancies on the electronic band structure was investigated. Monovacancies were created in a supercell containing $108$ atoms by removing each of the two symmetrically distinct oxygen sites  $6i$ and $2d$.  The resulting electronic pDOS for the two types of vacancies,  depicted in the two bottom panels of Figure~\ref{fig:DOS_hex_v1_v2_arrow}, shows that the vacancies create an occupied valence state in the band gap of the pristine HNS and essentially reduces the energy differences between conduction and valence bands. The values of the gaps are thus $2.1$~eV and $2.3$~eV for the $6i$ and $2d$ vacancies, respectively.

Another pathway to modify gaps in 2D materials is strain engineering.~\cite{conley_bandgap_2013} Since 2D sheets are the fundamental building blocks for layered heterostructures in functional materials, they are often grown through chemical vapor deposition (CVD) and physical epitaxy on a substrate or as mechanically assembled stacks.~\cite{Novoselov2016,Xu2013,Feng2016} Both approaches can lead to strain, and its effect on the band-gap was studied at the HSE06 level by applying three different kind of strains: biaxial strain simultaneously along the two equivalent lattice vectors $\vec{a}$ and $\vec{b}$, uniaxial strain along $\vec{a}$ only, and along $\frac{1}{2}\vec{a}+\vec{b}$. As shown in panel (b) of Figure~\ref{fig:eband_191_lepid}, the gap energies increase and decrease upon compressive and tensile strain, respectively, and the change only weakly depends on the direction of uniaxial strain. Although  a biaxial tensile strain of $10$\% leads to a large reduction of the band gap by 20\%, the gaps never reaches the range of visible light. Nevertheless, growing the HNS on an appropriate substrate would enable a continuous tuning of the gap in a wide range.

\begin{figure}
\includegraphics[width=0.9\columnwidth]{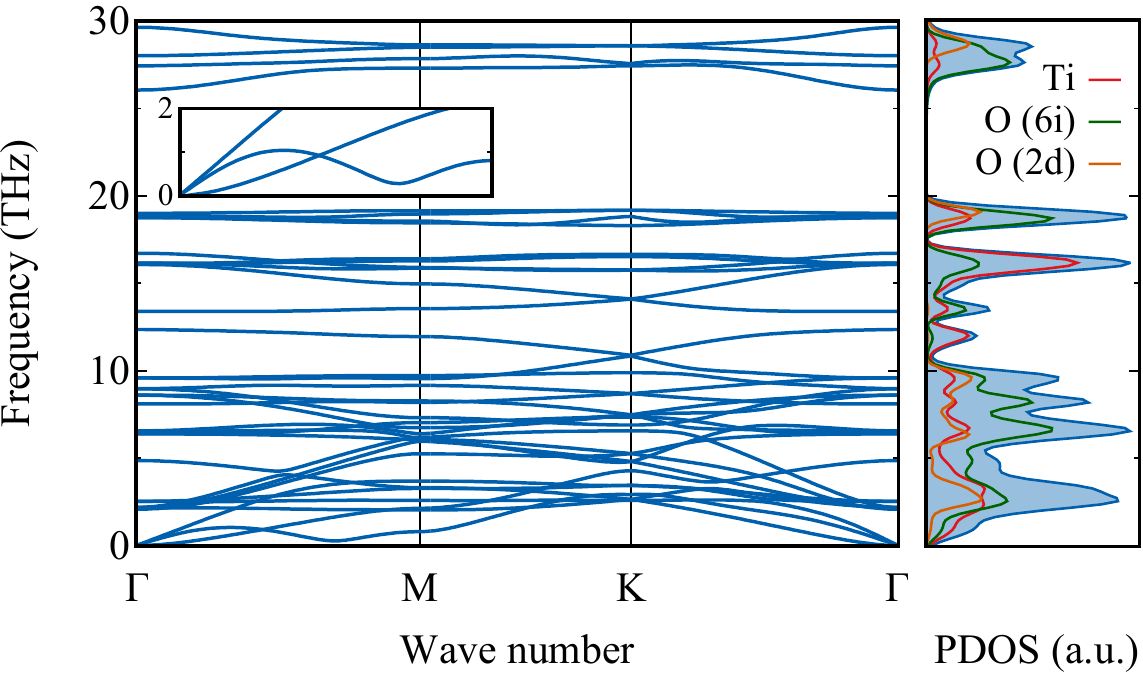}
\caption{The phonon band structure of HNS in the 2D Brillouin zone, together with the partial phonon density of states (PDOS). The shaded ares indicates the total density of states. The inset shows the acoustic branches along $\Gamma$-M, illustrating that the soft, acoustic phonon branch does not exhibit any imaginary modes. Furthermore, the dispersion clearly shows the two linear and one quadratic branches   as  M$\rightarrow\Gamma$, characteristic for 2D materials.
\label{fig:phonon}}
\end{figure}

Large strains experienced for example during the lithiation process in lithium ion batteries or hydrogenation in hydrogen storage applications can also compromise the mechanical stability of a layered material. Therefore, all internal degrees of freedom were fully relaxed when applying strain. Although the  atomic structure of the HNS remains intact even at $+10$\% tensile strain, the outermost layers of O atoms start to buckle at compressive strains below -$8$\% for uniaxial and below -$6$\% for biaxial strain, and the hexagonal voids start to deform slightly.
The dynamical stability of the HNS was assessed at its equilibrium lattice parameters by computing the phonon dispersion in the whole 2D Brillouin zone with the frozen phonon approach. The calculated phonon spectrum  in Figure~\ref{fig:phonon} shows no imaginary modes, indicating that the HNS is indeed viable and a metastable structures corresponding to a local minimum on the energy landscape. Like other 2D materials, two acoustic branches are linear as $q\rightarrow 0$.  Since force constants related to the transverse motion of the atoms decay rapidly, the lowest acoustic mode shows a quadratic behavior close to the $\Gamma$ point.~\cite{Cahangirov2009} In fact, this quadratic dispersion is common to all layered materials, as recently discussed in detail by Carrete~\textit{et al.}~\cite{carrete_physically_2016}

In addition to the mechanical properties, a high chemical stability is essential in order for a 2D material to be viable in practical applications. We can deduce a first evidence of the high chemical stability of the HNS from its band gap energy, which is the largest among all \ce{TiO2} polymorphs (Table~\ref{tab:energies}). Next, we analyzed the crystal orbital Hamilton population (COHP) and its integral (ICOHP) using the LOBSTER code.~\cite{Dronskowski1993,Deringer2011,Maintz2013,Maintz2016} 
Negative and positive values of the COHP correspond to bonding and antibonding states, respectively, while the ICOHP is considered here as a measure to compare bond strengths.~\cite{Maintz2013,Kishore2016} The  COHP between neighboring Ti and O atoms were calculated for the HNS, LNS and the anatase (001) slab. 
No antibonding interactions are present in the LNS and HNS for the occupied states. On the other hand, antibonding states exist for anatase (001) slightly below the Fermi level, indicating an electronic instability. Indeed, the anatase (001) undergoes a structural distortion and  transforms to the LNS sheet upon relaxation with a very tight force convergence criterion. 
Furthermore, a comparison of the -ICOHP at the Fermi level between the HNS and LNS results in values of $5.33$~eV and $3.54$~eV, respectively, indicating stronger Ti--O bonds in HNS, another evidence for its superior chemical stability.

\begin{figure}
\includegraphics[width=0.9\columnwidth]{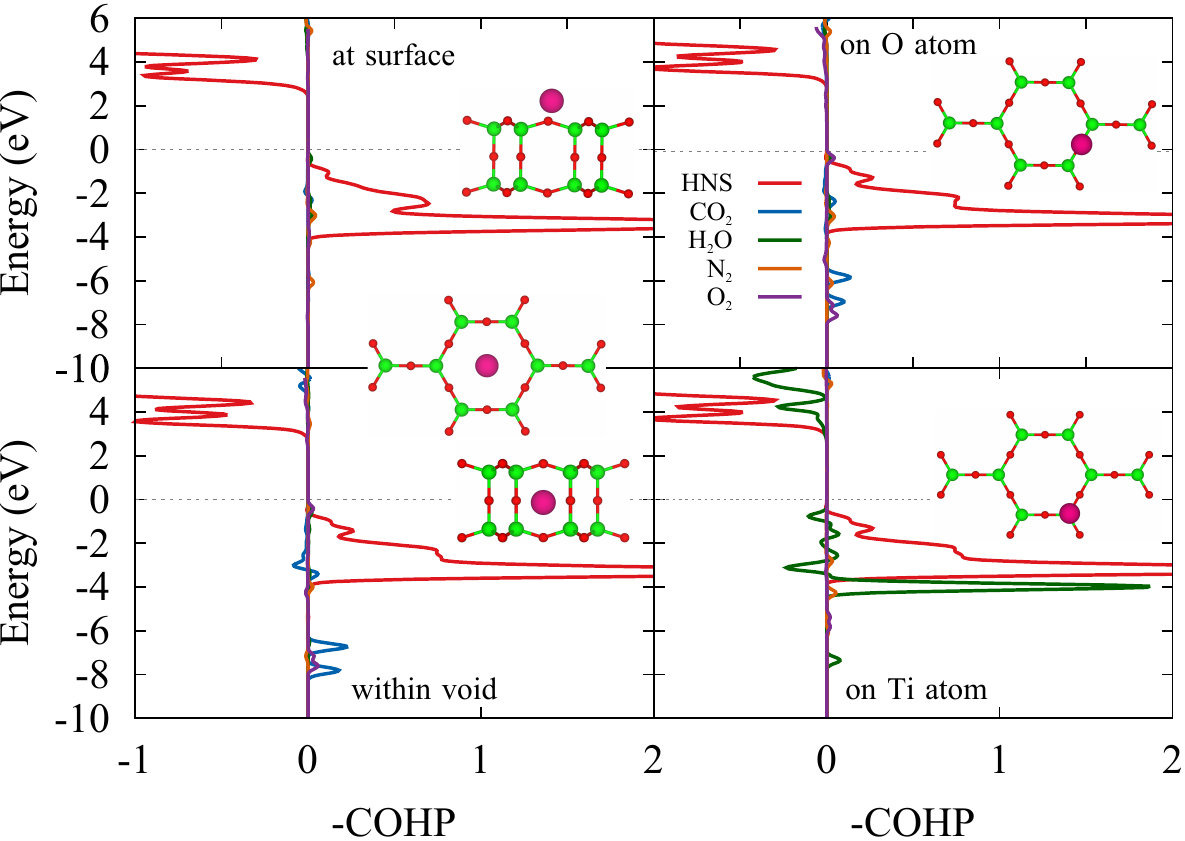}
\caption{COHP plots showing the Ti-O  bonding interaction of the HNS in red, together with the bonds between the HNS and \ce{CO2}, \ce{H2O}, \ce{N2} and \ce{O2} molecules. The geometric location where these molecules 
are adsorbed on the sheet is shown by the large, purple sphere in the insets.
\label{fig:COHP_molecule_191}}
\end{figure}

Furthermore, we verified the chemical stability of the HNS by investigating its potential reactivity with the most common molecules in the air, such as  N$_2$, O$_2$, CO$_2$, and H$_2$O.  For this purpose, these molecules were placed at different sites of the sheet and local geometry relaxations were performed to check if chemical bonds would form between the sheet and the molecules. Four different sites were investigated as shown in  Figure~\ref{fig:COHP_molecule_191}: on top of an O atom, Ti atom, and in the center of the hexagon at the surface of the sheet, and within the void of the hexagons. The initial distance from the sheet was chosen to be smaller than the sum of the covalent radii $r_\text{cov}^\text{s}$ of the closest two atoms of the adsorbate and the sheet whenever possible in order to induce a chemical reaction. However, all molecules placed on the surface were repelled immediately, leading to interatomic distances larger than $r_\text{cov}^\text{s}$. Similarly, molecules inside the sheet at the center of the hexagons remained intact and did not form chemical bonds with any atoms of the host structure. These findings were also supported by essentially no change in the charge density plots of the sheet and the adsorbate. A similar conclusion can be drawn from the COHP of the relevant atomic interactions, i.e. between the host and guest atoms closest to each other.  As shown in Figure~\ref{fig:COHP_molecule_191}, the interaction between the molecules and the HNS  is negligible for O$_2$, CO$_2$ and N$_2$. However, a relatively strong interaction is observed for the H$_2$O molecule placed on a Ti atom, which might be of interest in photocatalytic water-splitting application for hydrogen production.~\cite{ni_review_2007} Nevertheless, the magnitude of this interaction is much smaller compared to the Ti--O bonds within the sheet and we can safely conclude that even water will not disrupt the HNS structure.

If the HNS were to be grown or adsorbed on a substrate, 
its crystal and electronic structure will be affected due to the interaction at the interface. To study the effect of this interaction we deposited a layer of HNS on two substrates commonly used in CVD, namely Au~(111) and Ag~(111).
~\cite{Oznuluer2011,Kiraly2013}  All atoms were fully relaxed at the equilibrium lattice constant of the bulk material, and Van der Waals interactions were taken into account using the Tkatchenko-Scheffler method with Hirshfeld partitioning.~\cite{Tkatchenko2009} The atomic structure of the HNS remains overall unaffected during the relaxation, indicating that the sheet only weakly interacts with the substrates, leading to a final distance between the HNS and slabs of $2.9$~\AA~ for  Ag(111) and $3.1$~\AA~ for Au(111). These findings are supported by the COHP analysis, which show essentially no bonding states between Au--O/Ti and Ag--O/Ti, another strong indication that the HNS is chemically inert. We also investigated a heterostructure of a single HNS layer and a graphene sheet. Due to a small lattice mismatch, a supercell containing $16$~f.u. of \ce{TiO2} and $50$~C atoms is required to construct a commensurate cell, 
covering a very large surface area of $130.8$~\AA$^2$. The equilibrium distance between graphene and the HNS was found to be $3.22$~\AA. Again, a careful analysis of the COHP indicates that there are effectively no bonding states between HNS and graphene, providing the final evidence that the HNS is chemically inert.

In summary, we predict a novel 2-dimensional sheet of \ce{TiO2} by systematically searching for layered structures using an extended minima hopping structure prediction approach and \textit{ab initio} calculations. According to our results, this sheet is energetically favorable to any other previously reported  2D material of \ce{TiO2}, even surpassing the well known lepidocrocite sheet, and is dynamically stable with no imaginary modes in the whole Brillouin zone. A thorough study with respect to its structural and chemical stability shows that the HNS is chemically inert and highly resistant against a wide range of mechanical stress. In fact, uniform and uniaxial strain can be readily used to tune the band gap of the HNS. These findings provide strong evidences that the HNS is viable once synthesized and is a promising candidate material for strain engineered optical applications or for photocatalytic water-splitting due to its large surface area and voids. Furthermore, the HNS can readily serve as a building block for heterostructures with  potential applications in energy conversion and storage, ranging from lithium batteries to materials in photovoltaics. Due to its weak interaction with various substrates, we propose CVD as a potential route for its synthesis, or to use inert template atoms and thermal degassing to form the large hexagonal voids in analogy to the approaches used to create group-IV clathrate materials.

%*****************************************************************************************
\section*{Acknowledgments}
M.A. acknowledges support from the Novartis Universit{\"a}t Basel Excellence Scholarship for Life Sciences and the Swiss National Science Foundation (P300P2-158407). This work was partially performed within the Swiss National Competence Center for Research,  Marvel. We gratefully acknowledge the computing resources from the Swiss National Supercomputing Center in Lugano (project s700 and s707), the Extreme Science and Engineering Discovery Environment (XSEDE) (which is supported by National Science Foundation grant number OCI-1053575), the Bridges system at the Pittsburgh Supercomputing Center (PSC) (which is supported by NSF award number ACI-1445606), the Quest high performance computing facility at Northwestern University, and the National Energy Research Scientific Computing Center (DOE: DE-AC02-05CH11231).

\section*{Methods}

The structural search was conducted with the minima hopping method (MHM),~\cite{Goedecker2004,Amsler2010} which implements an efficient algorithm to explore the energy landscape of a system by performing consecutive molecular dynamics (MD) escape steps followed by local geometry relaxations. The Bell-Evans-Polanyi principle is exploited to accelerate the search by aligning the initial MD velocities preferably along soft mode directions.~\cite{Jensen} In order to perform predictions for quasi-2D materials the functionality of the MHM was  modified to model 2D and layered materials. For this purpose, the target function to be optimized was extended by including 2-dimensional confining potentials $C_i(e_i,\textbf{r}_j^\alpha)$, where $\alpha$ denotes the axis $\alpha = \{x,y,z\}$ along the non-periodic direction, $\textbf{r}_j$ are the cartesian coordinates  of the $N$ atoms in the system, and $e_i$ are the equilibrium positions along $\alpha$ at which the potentials are centered. Our confinement functions are sums of atomic contributions and is zero within a cutoff region $r_c$ around $e_i$, while it has a polynomial form of order $n$ with amplitude $A_i$ beyond $r_c$: $C_i^\alpha=\sum_{j=1}^N c(e_i,\textbf{r}_j^\alpha)$ where $c(e_i,\textbf{r}_j^\alpha)=A_i(|e_i - \textbf{r}_j^\alpha|-r_c)^n$ for $|e_i-\textbf{r}_j^\alpha|\geq r_c$  and zero otherwise. During the local geometry optimization and the MD escape trials the additional forces acting on the atoms $\textbf{f}_j=\frac{\partial C_i}{\partial \textbf{r}_j}$ and on the cell vectors $\sigma_j=\frac{\partial C_i}{\partial h_j}$ were fully taken into account, where the atomic positions are expressed in the reduced coordinates $\textbf{r}_i=h\textbf{s}_i$ and $h=(\textbf{a}, \textbf{b}, \textbf{c})$ is the matrix containing the lattice vectors. For our calculations, a single confinement potential $C_1$ was used with $e_1$ centered along the lattice vector $\textbf{c}$, with a cutoff $r_c=1$~\AA, $n=4$ and $A_1=0.1$~eV. Multiple MHM structure prediction runs were performed, starting from a variety of random initial seeds using 2-10~f.u./cell in conjunction with the CENT potential.

To accelerate the structural search, we prescreened the energy landscape using a recently developed high dimensional artificial neural network (ANN) potential based on the charge equilibrated neural network technique (CENT)~\cite{Ghasemi2015} specifically fit to the \ce{TiO2} system. In order to train the CENT potential we first prepared a database containing the DFT energies of about $3000$ cluster structures, with sizes ranging from $6$ to $70$ formula units (f.u.). An approximate potential was obtained by fitting CENT to DFT data of randomly generated structures in an initial step. Subsequently, further structures were added to the training set by performing short MHM simulations with the approximate potential. The resulting low energy structures from these MHM runs were carefully filtered to avoid duplicate structures and to ensure a large diversity in structural motifs within the augmented data set, a task performed by comparing structures with a structural fingerprint~\cite{fingerprint_2016,sadeghi_metrics_2013}. This procedure was repeated several times until a desired accuracy was reached for a predefined validation dataset. Even though the CENT potential was generated from cluster structures, it provides accurate results for periodic systems. This high transferability of CENT was recently demonstrated for calcium fluoride.~\cite{faraji_high_2017}

The DFT calculations to generate the training data and to refine the results from the MHM simulations were carried out with the  FHI-aims code,~\cite{Blum2009} using the tier 2 basis set for both the  Ti and O elements. The generalized gradient approximation parametrized according to  Perdew-Burke-Ernzerhof (PBE)~\cite{Perdew1996} was employed for the exchange-correlation functional. Geometry relaxations were performed until the atomic forces were less than $0.01$~eV/\AA, and Monkhorst-Pack k-point grids with a density of $0.03/$\AA~ were used, resulting in total energies converged to within 
$1$~meV/atom. For all calculations involving 2D structures a vacuum space of about $12$~\AA~ was used in the direction perpendicular to the sheets to suppress the interaction between periodic images of the layers.

The chemical bonding was studied by analyzing the crystal orbital Hamilton population (COHP) as implemented in the LOBSTER package,~\cite{Dronskowski1993,Deringer2011,Maintz2013,Maintz2016} based on the wave functions computed with the Vienna Ab Initio Simulation package (VASP).~\cite{Kresse1993} Phonon calculations were carried with the finite difference approach as implemented in the PHONOPY code,~\cite{Togo2015} using atomic displacements of  $0.01$~\AA. The dielectric tensor and the born effective charges were computed with VASP. Large supercells containing $192$ atoms were used to ensure sufficient  convergence of the force constants.

%*****************************************************************************************
% \bibliography{tiO2layer}% Produces the bibliography via BibTeX.
%merlin.mbs apsrev4-1.bst 2010-07-25 4.21a (PWD, AO, DPC) hacked
%Control: key (0)
%Control: author (8) initials jnrlst
%Control: editor formatted (1) identically to author
%Control: production of article title (-1) disabled
%Control: page (0) single
%Control: year (1) truncated
%Control: production of eprint (0) enabled
\providecommand{\noopsort}[1]{}\providecommand{\singleletter}[1]{#1}%

\end{document}